# Quantifying firm-level risks from nature deterioration

Ricardo Crisóstomo[*]

## Abstract

We estimate the loss of value that companies might suffer from nature overexploitation. We find that global equities shed 26.8% in a scenario of unabated nature decline, while the worst-performing firms lose ~75% of their value. Our risk framework considers five environmental hazards: biodiversity loss, land degradation, climate change, human population and nature capital. We also introduce two metrics to assess nature-related risks: a Country Degradation Index that tracks the damage caused by environmental hazards in specific territories, including nonlinear dynamics and tipping points; and a Nature Risk Score that summarizes the risk that companies face due to the decline of nature and its services.

[*]Comisión Nacional del Mercado de Valores (CNMV), Edison 4, 28006 Madrid. The author acknowledges useful comments and assistance from Beatriz Boixo.
Email: rcayala@cnmv.es

# 1. Introduction

Human life quality has experienced extraordinary growth over the last century. Global life expectancy has increased from 36.4 to 73.2 years, extreme poverty has fallen from 32% to 7%, and child mortality has decreased sixfold[2]. However, the key role that nature has played in our prosperity has been largely unnoticed. Every day, we rely on natural resources and services that sustain human life. WEF (2020) estimates that over half of the world's GDP is moderately or highly dependent on nature. Similarly, Costanza et al. (2014) estimate that the annual value of ecosystem services amounts to US$125 trillion, about 1.5 times the global GDP.

The availability of vital goods and services —such as freshwater, food, clean air or fertile soils— requires a well-functioning natural world. The importance of ecosystem services is such that humanity, and our global economy, could not exist without them. However, since 1970 our demand for natural resources has exceeded the rate at which the biosphere can regenerate them. As a result, nature is deteriorating and many of the essential services on which life depends are degrading at unprecedented rates[3].

The current trend of nature overexploitation poses stark economic risks. Scientists warn that if we do not halt the loss of nature and its services, even the Earth's habitability could be compromised. Consequently, failure to account for, mitigate, and adapt to nature-related risks creates huge social and financial stability risks, bringing environmental degradation within the scope of financial supervisors[4].

Measuring nature-related risks is, however, a complex task. The processes and systems that underpin environmental degradation are multifaceted, nonlinear and characterized by tipping points[5]. Consequently, nature risk assessments show a disturbing uncertainty about the likelihood and consequences of potentially catastrophic events. But given the risks at stake, uncertainty should not be an excuse for inaction. Since there is no planet B, a precautionary approach that limits the risk of irreversible disruptions is advocated in Earth system analyses.

This paper contributes to the literature by quantifying the loss of value that companies could suffer from nature degradation. We introduce a Nature Risk Score ($NRS$) that assesses the overall risk that firms face due to the decline of nature and its services. The $NRS$ employs a comprehensive risk framework that jointly considers: (i) projections of five environmental hazards: biodiversity loss, land degradation, climate change, population growth and nature capital evolution; (ii) an assessment of firms' vulnerabilities to nature deterioration and (iii) an analysis of firms' exposures in specific territories.

---

[2] See Zijdeman and de Silva (2014), OWID (2023), Volk and Atkinson (2013) and https://ourworldindata.org.
[3] See IPBES (2019), Dasgupta (2021) and data.footprintnetwork.org.
[4] See NGFS-INSPIRE (2021), IPCC (2023) and WEF (2024).
[5] See Kedward et al. (2020), Lenton et al. (2019), Rockström et al. (2009) and Svartzman et al. (2022).

We employ two methods to estimate the financial consequences of nature risk. First, we analyze the loss of value that firms could suffer in the stock market given their environmental and financial risk profile. In addition, we estimate firms' losses through a discounted cash flow valuation where future cash flows are impacted by nature risk.

By combining country-level projections of nature degradation, sectoral vulnerabilities and firms' geographical exposures, we find that nature overexploitation poses a significant threat to our environment and the global economy:

- **Nature loss**: We estimate that 53% of the world's natural resources and services could be damaged or impaired by mid-century. For 174 countries representing 93% of global GDP, the damages projected for ecosystems range from 14% to 79%. The impact of nature deterioration is particularly acute in Sub-Saharan Africa, Europe and North America.

- **Vulnerability to nature**: We find that all economic sectors exhibit significant nature-related dependencies. Across 87 industries, the vulnerability to nature varies from 38% to 95%. Companies operating in agriculture, utilities and transport exhibit the largest nature-related risk given their dependencies on water-related services, pollination or climate regulation, among others.

- **Firm losses**: We conclude that nature degradation creates huge financial risks for companies. Based on 1454 MSCI World firms, we estimate that global equities could suffer an average 26.8% loss in a business-as-usual scenario of continued nature decline. However, the nature risk distribution is notably skewed, with the worst-performing firms shedding ~75% of their value.

The remainder of the paper is structured as follows: Section 2 presents our nature risk framework. Section 3 explains the data and methods used in the risk calculations. Section 4 shows the results of our analyses for a wide range of countries (174), business activities (601) and individual firms (1454). Finally, section 5 concludes.

## 2. Framework to measure nature-related financial risks

Measuring nature-related financial risks is an emerging field. There are still few studies that assess the implications of nature degradation for financial portfolios. The pioneering work of van Toor et al. (2020), analyze the dependencies of Dutch financial institutions to ecosystem services, finding that 36% of their portfolio is moderately or highly exposed to nature. Svartzman et al. (2022) employ a related methodology, showing that 42% of French portfolios are dependent on nature. Similar studies (Calice et al., 2021; World Bank, 2022; Boldrini et al., 2023) report dependencies ranging from 46% to 72%, concluding that financial portfolios and institutions are significantly exposed to nature.

We contribute to the literature by developing a comprehensive framework to assess the overall risk that companies face from nature deterioration[6]. In line with Svartzman et al. (2022), our framework integrates three risk components: (i) a scenario describing the evolution of multiple environmental hazards (ii) an analysis of firms' exposures to these natural hazards and (iii) an evaluation of firms' propensity to face losses given the projected scenario and their nature-related exposures. Figure 1 provides a schematic description of the steps and risk factors used in our nature risk framework[7].

**Figure 1: Overview of our nature risk framework**

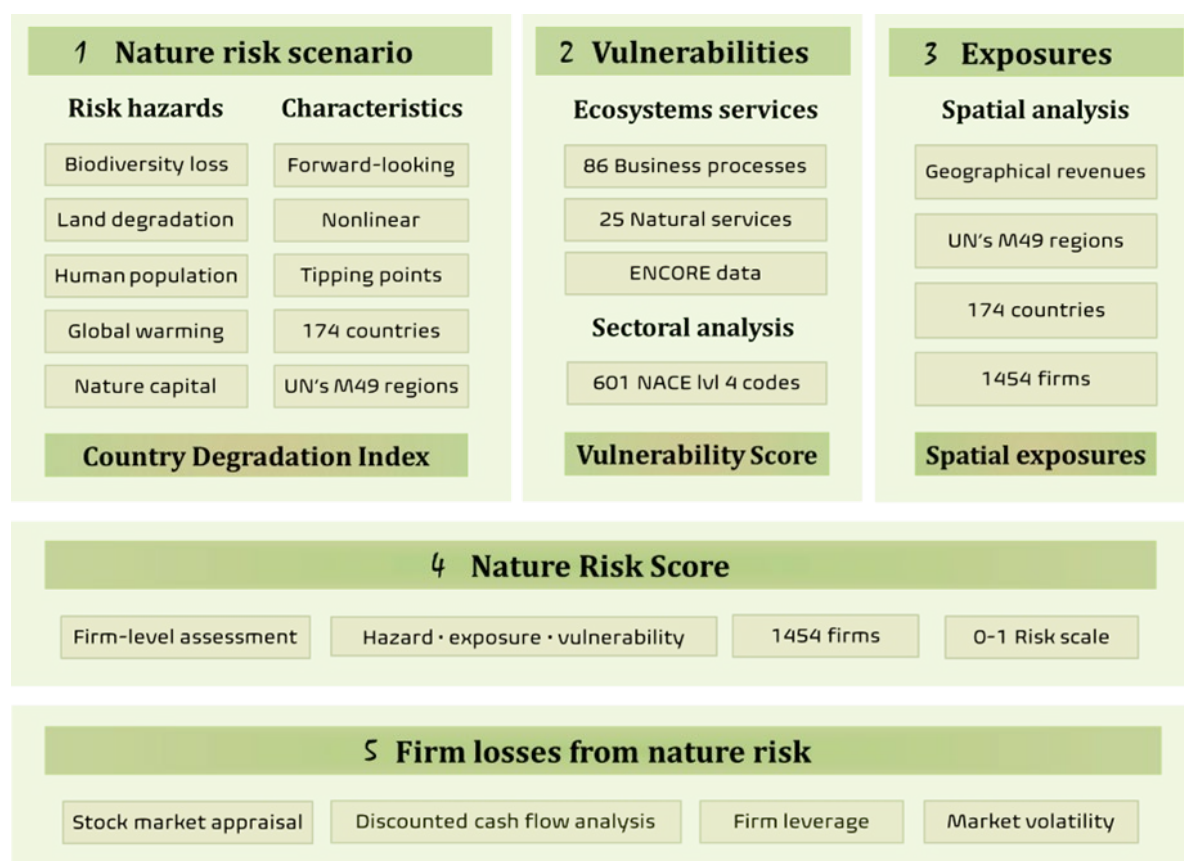

---

[6] As indicated in ECB/ESRB (2023) and NGFS (2023), there are still no full-fledged analyses that quantify the risks that companies face due to nature degradation.
[7] Upon completion of our risk framework, Ranger et al. (2023) published a scenario-based analysis that employs a comprehensive hazard/exposure/vulnerability assessment to measure nature-related risks.

## 2.1 Nature risk scenario

Our risk scenario considers nature-related shocks that are multidimensional, nonlinear and forward-looking. Environmental degradation is characterized by a myriad of interrelated factors that cannot be summarized by a single metric. In addition, growing environmental pressures may trigger nonlinear dynamics and tipping points, requiring modelling frameworks that account for these complexities. Furthermore, since we are increasingly moving into uncharted territory in nature overuse, nature risk scenarios should consider the materialization of future events that differ from those in historical data.

Specifically, our nature-risk scenario considers five key environmental hazards:

**Biodiversity loss.** Biodiversity is the variety of life on Earth. Without biodiversity, nature cannot provide the essential goods and services that sustain human life. Despite being indispensable for food security, air filtration, clean water provision, disease prevention, and many other services, biodiversity is declining faster than ever. The World Economic Forum ranks biodiversity loss as the third most impactful risk for the global economy.[8]

**Land degradation.** Land degradation is the loss of biological and economic productivity of land. Each year, land degradation costs 6.3 trillion US dollars (7.3% of global GDP); and with food demand expected to increase over 50% by 2050, access to fertile soils emerges as a key driver of food security. Degraded land also reduces freshwater quality and availability, and increases vulnerability to environmental disasters (e.g., floods)[9].

**Global warming.** Rising temperatures, changing weather patterns and weather extremes pose a significant threat to ecosystems. Warmer temperatures alter natural habitats by changing what can grow and live within them[10]. Similarly, weather extremes destroy ecosystems, reducing their capacity to provide essential goods and services. Furthermore, global warming and nature loss are mutually reinforcing; as weather extremes destroy wild habitats, ecosystems release carbon and reduce their capacity to act as carbon sinks, increasing $CO_2$ concentrations, which intensify climate change and fuel more weather extremes[11].

**Population growth.** We can seldom halt —let alone reverse— the loss of nature if we continue growing, consuming, producing and polluting at the current pace.[12] Human appropriation of Earth's biosphere is, arguably, the main systemic factor behind the decline of natural ecosystems. Scientists warn that human activity has already pushed our planet beyond safe operating limits in six of nine critical boundaries, damaging its ability to self-regulate and increasing the risk of catastrophic events (Richardson et al., 2023; Rockström et al., 2009).

---

[8] See CISL (2020), WEF (2020, 2024) and IPBES (2019).
[9] See CISL (2020), Sutton et al. (2016) and Valin et al. (2014), UNCCD (2011) and CISL (2021).
[10] For example, higher temperatures and ocean acidification are severely impacting corals, which have declined by half since 1950 (IPCC, 2021, Eddy et al., 2021 and Svartzman et al., 2022).
[11] See Pörtner et al. (2023) and NGFS (2023).
[12] We are using the equivalent of 1.6 Earths to maintain our current way of life, and ecosystems cannot keep up with our demands. See UNEP (2021) and https://data.footprintnetwork.org/

**Nature capital depletion:** Natural capital is the stock of natural resources, which includes water, soils, air, and living organisms. Destruction and overexploitation of natural resources threaten the availability of the essential services on which life depends. Countries deprived of natural resources may face food scarcity, water stress, high prices of raw materials and social unrest (e.g., armed conflict and mass migration). The loss of nature capital is also linked to a deterioration of human health and wellbeing (Lu & Sohail, 2022).

To model the evolution of these risk hazards, we project their trajectories in a business-as-usual scenario of unabated nature decline. Next, to quantify the damages inflicted by nature degradation, we consider both nonlinear dynamics and tipping points:

- **Nonlinear effects.** Environmental damages progressively accelerate as the loss of nature intensifies. Nonlinear impacts are justified by the limited availability and low substitutability of natural resources. Intuitively, nature-related benefits like clean air, freshwater availability or fertile soils cannot be easily substituted, making them more valuable as they are progressively lost. Similarly, the impact of losing an essential service (e.g., pollination) is greater when the service is already scarce than when it is abundant[13].

- **Tipping points.** As nature is progressively lost, there is also an increasing risk of crossing an ecological threshold, triggering changes in Earth's biosphere that lead to irreversible effects. Ecological boundaries are broadly recognized in Earth-system science[14]. However, there is still a notable uncertainty about the likelihood and consequences of crossing a tipping point. Consequently, we model ecological thresholds using a probabilistic approach.

To summarize the cumulative impact of all risk hazards, including nonlinear effects and tipping points, we introduce the **Country Degradation Index ($CDI$).** The $CDI$ quantifies the expected damages from nature loss in specific countries and is used along with firms' exposures and vulnerabilities to evaluate the risks that companies face due to the decline of nature and its services.

## 2.2 Geographical exposures

We assume that firms' exposures to environmental shocks are mainly driven by location. Firms with business or assets in specific regions are particularly exposed to the deterioration of nature in these regions. For example, the effects of water stress, air pollution and natural disasters are primarily felt in the areas where they occur.

To evaluate geographical exposures, we use the spatial distribution of firms' revenues. The distribution of revenue provides a readily available proxy to assess firms' interests in particular regions. Alternatively, we acknowledge that asset-level data could provide a higher level of granularity. However, asset-level information is limited or unavailable for most companies in our dataset, and we could not obtain this information.

---

[13] See Kedward et al. (2020) and Svartzman et al. (2022).
[14] See Lenton et al. (2019) and Armstrong McKay et al. (2022).

## 2.3 Vulnerabilities to nature loss

Beyond geographical exposure, firms' vulnerabilities to nature degradation vary by economic sector. Companies operating in specific sectors have distinct dependencies on environmental services. For example, agriculture firms depend on soil formation to maintain fertile land and pollination to enhance crop yields. In contrast, manufacturing firms depend on genetic diversity and the availability of raw materials to produce new and existing products.

To obtain nature-related vulnerabilities, we employ the Ecosystem Services Framework (Turner & Daily, 2008). Specifically, we rely on ENCORE[15] to analyze the dependencies of firms' production processes on a broad set of ecosystem services, quantifying companies' vulnerabilities to natural services such as water availability, climate regulation or provision of raw materials, among others.

## 2.4 Nature Risk Score

The $NRS$ employs a *hazard · exposure · vulnerability* framework to perform a comprehensive nature risk assessment. Specifically, the $NRS$ classifies each firm on a 0–1 risk scale by aggregating: (i) a forward-looking scenario of nature-related shocks [*hazards*]; (ii) firms' exposures in specific countries [*exposures*] and (iii) firms' dependencies to the loss of nature and ecosystem services [*vulnerabilities*]. By combining these risk components, the $NRS$ performs a full-fledged analysis that measures the overall risk that companies face from nature degradation.

## 2.5 Firms' losses from nature risk

Assessing the economic impact of nature deterioration is essential for integrating nature risks into decision-making. However, measuring the consequences of nature loss is challenging due to the complex, multidimensional, nonlinear and interacting factors that characterize environmental degradation. Despite recent advances, there is still no consensus on how to value nature and its services, calling for additional research to assess the financial impact of nature loss (ECB & ESRB, 2023; IPBES, 2022; NGFS, 2023).

Given the lack of standards, our risk framework employs two complementary methods to quantify the economic losses associated with nature risk. In a scenario of continued nature decline, firms with large nature-related exposures and vulnerabilities will face higher future costs and lower revenues. Hence, to model the link between nature risk and financial stress, we first employ a simple model where the $NRS$ of each firm, along with its volatility and financial leverage, is used to estimate stock returns. In addition, we estimate firms' losses through a discounted cash flow valuation where future cash flows are impacted by nature risk.

---

[15] ENCORE (Exploring Natural Capital Opportunities, Risks and Exposures) is an online tool that assesses how different sectors, subsectors and production processes depend and impact on nature. See https://encorenature.org/

## 3. Data and methodology

This section presents the data and methods used to quantify firms' losses due to nature degradation. Our framework integrates data on environmental hazards, firm-specific exposures, and sectoral vulnerabilities to provide a comprehensive assessment of nature-related risks.

### 3.1 Environmental hazards

Table 1 summarizes the environmental data used in our nature risk scenario. Since nature degradation is driven by local and systemic factors, we consider risk hazards with different granularities, from global to country-level, to model the accumulation of environmental pressures[16].

Table 1: Environmental hazards data

| Environmental hazard | Data Source | Projection method | Spatial granularity | |
|---|---|---|---|---|
| Biodiversity loss | IUCN Red List Index | Grade 3 polynomials | Subregional | [10] |
| Land degradation | UN's statistics | Linear estimates | Country-level | [174] |
| Global warming | IPCC / NGFS | Forward-looking data | Continental | [6] |
| Human population | UN's statistics | Forward-looking data | Global | [1] |
| Natural capital | Nature Capital Index | Linear estimates | Country-level | [174] |

Notes: UN's M49 geographical standards are employed to define world regions.

**Biodiversity loss.** Biodiversity projections are based on the IUCN Red List Index (RLI). The RLI tracks the state of biodiversity by assessing the extinction risk of species. Each species is rated on a 0–1 risk scale, with 1 representing no risk of extinction and 0 an already extinct species. By assessing the extinction of species over time, the RLI provides a time series of biodiversity that has been widely used to track biodiversity goals[17].

We retrieve RLI data from the advanced search function of the IUCN webpage. Specifically, we download the RLI for 10 world regions in the 2000-2020 period[18]. Next, to obtain forward-looking estimates, we compute the year-on-year (YoY) changes on each regional index, and biodiversity hazards are projected using grade 3 polynomials.

**Land degradation:** Despite being a severe threat, quantifying land degradation has proven difficult.[19] To ensure data consistency, we employ the dataset curated by the United Nations (UN) Statistics Division[20]. UN's data provides the percentage of degraded land for 174 countries in 2015 and 2019. To construct our scenario, we calculate the proportion of degraded land in the sample period and, owing to the scarcity of data points, we project country-level trends using linear estimates.

---

[16] See NGFS (2023).
[17] For instance, the RLI has been used to track progress towards the UN's Sustainable Development Goals (SDGs) and employed by the post-2020 Global Biodiversity Framework. See https://www.iucnredlist.org
[18] The list of IUCN subregions is available upon request.
[19] Diverging views on how to define degraded lands and calculation methods have led to controversy (Gibbs & Salmon, 2015; IPCC, 2019; Prince et al., 2018; van der Esch et al., 2017).
[20] UN's data has been used to track progress towards SDG goal 15.3.1. See https://unstats.un.org/sdgs

**Global warming.** Forward-looking temperature paths are retrieved from the Climate Impact Explorer of the NGFS[21]. The NGFS analyses several global warming scenarios. One scenario considers that only currently implemented policies are preserved, following the narrative of a business-as-usual scenario where no additional efforts are made to mitigate climate change. Although climate change is a global phenomenon, the impact of rising temperatures is felt differently across territories. Hence, to discriminate by region, we retrieve temperature projections for the six continents included in the NGFS database.

**Human population.** We obtain forward-looking population projections from the UN's population division. Human appropriation of Earth's biosphere has already caused significant environmental degradation, threatening the life-supporting services that nature provides. Looking ahead, a growing population following a business-as-usual lifestyle will be a key driver of nature loss. Therefore, to model human-induced pressures, we retrieve global population projections and human growth is used to assess the anthropogenic forcing driving nature loss across different regions.

**Natural capital depletion:** Trends in natural capital are retrieved from the Natural Capital Index (NCI)[22]. The NCI assesses the state of nature capital through a broad set of environmental indicators, including water quality and availability, pollution and mineral resources, among others. To obtain consistent series, we retrieve the NCI for 174 countries from 2018 to 2022[23]. Next, we calculate the YoY changes for each country and, given the volatility of NCI scores, forward-looking forecasts of nature capital are obtained using linear estimates.

### 3.2 Country Degradation Index

To quantify the impacts of nature deterioration, we first consider the pressures driven by environmental hazards in specific countries. Next, we estimate the damages caused by environmental pressures using impact functions that include nonlinear effects and tipping points. Finally, we aggregate the impacts from all risk hazards through a country degradation index that tracks the damages caused by nature deterioration in specific countries.

### 3.2.1 Environmental pressures

Denoting $\lambda_{t,c}^k$ the intensity of environmental hazard $k$, in year $t$ and country $c$, the cumulative pressure driven by hazard $k$ is obtained as

$$EP_{t,c}^k = 1 - e^{\sum_{t=t_0}^{T} -\lambda_{t,c}^k} \quad (1)$$

For the risk hazards that are directly expressed as nature loss (i.e. biodiversity loss, land degradation and natural capital depletion) equation (1) tracks the build-up of environmental pressures across time. For instance, if the land degradation pressure is 0.2 in country $c$, it indicates that 20% of the land has degraded from $t_0$ to $T$.

---

[21] Global mean temperature projections are sourced from the REMIND-MAgPIE and MAGICC climate models. See https://climate-impact-explorer.climateanalytics.org
[22] The NCI is developed by SolAbility.
[23] Due to methodological changes, we employ only recent readings of the NCI to ensure data consistency.

Alternatively, to assess the pressures driven by climate change and population growth, an upper threshold is required to map the accumulation of risk hazards. For global warming, a global mean surface temperature increase of ~1.1°C has already led to widespread impacts[24]. Scientists predict that exceeding 1.5°C above pre-industrial levels will substantially increase harm to nature and people, threatening the Earth's resilience and stability[25]. Consequently, we consider an increase of 3°C in 2050 as the threshold to map climate change pressures. This level of warming is higher than the worst-case scenario considered by the IPCC[26] and twice the 1.5°C included in the Paris Agreement.

Similarly, with a population of close to 8 billion, we are consuming Earth's resources much faster than they can regenerate[27]. Human activities have already pushed the Earth beyond six of nine planetary boundaries, threatening nature's life-supporting services. A growing human population is expected to further accelerate the rate of nature destruction, increasing the risk of catastrophic impacts. Consequently, we consider an additional increase of 50% in human population in 2050 (up to ~12 billion) as the threshold to map human growth pressures[28].

**3.2.2 Environmental damages**

To quantify the damages caused by environmental pressures, we consider both nonlinear effects and tipping points.

**Nonlinear dynamics.** Following Keen (2023) and Trust et al. (2023), we employ a logistic function to model environmental damages. Specifically, we model the harm caused by growing environmental pressures as:

$$d_{t,c}^k = EP_{t,c}^k + \frac{1 - EP_{t,c}^k}{1 + e^{-10(EP_{t,c}^k - 0.5)}} \qquad (2)$$

Where $d_{t,c}^k$ represents the damage expected from environmental pressure $k$. Consequently, our damage function aggregates two components: (i) an impact proportional to the accumulation of risk pressures and (ii) a nonlinear impact that measures the acceleration of damages as nature is progressively lost. Combining both components, we obtain a nonlinear function that quantifies the harm caused by environmental hazards on a 0–1 scale[29].

---

[24] See IPCC (2023)
[25] See Armstrong McKay et al., 2022; OECD, 2022; Rockström et al., 2009; Xu & Ramanathan, 2017.
[26] The most extreme SSP5-8.5 IPCC AR6 scenario projects an average temperature increase of ~2.5°C in 2050.
[27] See UNEP (2021). We are using the equivalent of 1.6 Earths to maintain our current way of life.
[28] Given the lack of accepted thresholds, we consider that a global population of ~12 billion and a warming level of 3°C in $T$ = 2050 represent a reasonable proxy of a future ruin scenario that can be used to reverse engineer the risk pressures at time $t$, as proposed in Trust et al. (2023). Specifically, we consider the thresholds at $T$ as the upper level of the 0–1 pressure scale, obtaining the $t$-pressures by linearly interpolating from the initial values at $t_o$ (end-2022) and the upper thresholds at $T$.
[29] Environmental damages are generally modelled with a lower than linear impact for small risk pressures. However, world's nature at end-2022 is already quite damaged and far from pristine. Hence, given the current state nature deterioration, we employ an initial linear impact that progressively accelerates as environmental pressures increase (see e.g., Scheffer et al, 2001; Estes et al., 2011; Newbold et al., 2020; and Acheampong & Opoku, 2023).

Next, to incorporate the interdependences across risk hazards, we estimate the average damage expected in each country as:

$$\bar{d}_{t,c} = \frac{\sum_{k=1}^{n} d_{t,c}^{k}}{n} \qquad (3)$$

By aggregating the impact of different risk hazards, $\bar{d}_{t,c}$ provides a comprehensive metric that can be used to track nature degradation in specific territories, and gauge the interactions and feedback loops that arise when environmental pressures accumulate in a given region.

**Tipping points**. To complement the gradual damage in equations (2) and (3), we also consider that growing environmental pressures can lead to trespassing an ecological threshold[30]. Following Keen (2023) and Trust et al. (2023), we model the probability of crossing a tipping point as:

$$p_{t,c} = \frac{1 - \bar{d}_{t,c}}{1 + e^{-10(\bar{d}_{t,c} - 0.5)}} \qquad (4)$$

Hence, in our risk framework, environmental tipping points are modeled probabilistically as a function of accumulating environmental pressures, as measured by $\bar{d}_{t,c}$[31]. To calibrate the impact of ecological thresholds, we consider the review by Dietz et al. (2021). In a recent study, Dietz et al. (2021) quantify the effect of crossing eight tipping points, measuring their economic consequences and their impact on nature and human health. Compared to an Earth-system without ecological thresholds, the materialization of tipping points increases the expected damage by 28.9%. Hence, we consider that tipping points, when triggered, increase environmental damages by a factor $\pi$ = 28.9%, as reflected in equation (5)[32].

$$d_{t,c}^{tp} = (1 - \bar{d}_{t,c}) \pi \qquad (5)$$

### 3.2.3 Country Degradation Index

The $CDI$ summarizes the damages expected from nature-risk hazards in specific countries. For each date $t$, the $CDI$ is obtained as:

$$CDI_{t,c} = \bar{d}_{t,c} + p_{t,c} d_{t.c}^{tp} \qquad (6)$$

Consequently, the $CDI$ aggregates the nonlinear impacts from all environmental pressures, $\bar{d}_{t,c}$, and the expected damage from tipping points, $d_{t.c}^{TP} p_{t,c}$. For each country and date, the $CDI$ tracks the extent of nature degradation in a 0–1 scale. A $CDI_{t,c}$ of 0 represents no damage from environmental hazards in country $c$, whereas a $CDI_{t,c}$ of 1 implies a complete destruction of nature and its services.

---

[30] See Armstrong McKay et al.(2022), Botero et al. (2015); Hillebrand et al. (2020) and Richardson et al. (2023).
[31] Using a probabilistic approach allows a straightforward modelling of tipping points, circumventing the need to set specific trigger levels and the uncertainty about how to characterize tipping points compounding and cascading effects.
[32] Specifically, to incorporate the impact of tipping point in a 0–1 scale, we consider that a proportion $\pi$ of the nature that has not been damaged $(1 - \bar{d}_{t,c})$ is impaired when a tipping point is crossed.

## 3.3 Firms' exposures and vulnerabilities

Consistent with the granularity of nature risk hazards, we obtain firms' geographical exposures in 174 countries. For each firm, the exposure to country $c$ is obtained as:

$$Exp_c^i = \frac{Rev_c^i}{TotalRev^i} \quad (7)$$

where $Rev_c^i$ is the spatial revenue of firm $i$ in country $c$, and $TotalRev^i$ the total revenue of firm $i$.[33] When companies' reported data include regional aggregates, country-level exposures are obtained through their GDP shares within the region.

Nature-related vulnerabilities are derived from ENCORE. ENCORE calculates the dependencies of 86 production processes on 25 ecosystem services, assigning materiality ratings ranging from zero to very high. Following van Toor et al. (2020) and Svartzman et al. (2022), we map ENCORE's materiality ratings to numerical scores, employing increments of 0.2[34].

To calculate firms' nature vulnerabilities, we first consider the production processes associated with each NACE level 4 code[35]. Next, we obtain the ecosystem dependencies of each production process, and we calculate the average of the highest nature-related dependencies across the firm-specific production processes. Hence, the vulnerability score to nature, $VS^i$, is given by:

$$VS^i = \frac{\sum_{h=1}^{n} ND^h}{n} \quad (8)$$

Where $h$ represents the number of production processes associated with NACE level 4 code, and $ND^h$ is the highest nature-related dependency of each production process.[36]

## 3.4 Nature Risk Score

The $NRS$ aggregates in a single measure the deterioration of nature expected in specific countries, firms' exposures to these countries, and firms' vulnerabilities to nature loss. Specifically, we calculate the risk that each company faces from nature deterioration using a risk = *hazard · exposure · vulnerability* framework, where

$$NRS^i = \sum_c CDI_{T,c} \cdot Exp_c^i \cdot VS^i \quad (9)$$

---

[33] Spatial revenues are retrieved from Refinitiv at end-2022. For companies where Refinitiv does not provide geographical data, Bloomberg is employed as an alternative.
[34] ENCORE assigns six materiality ratings (no dependency, very low, low, medium, high and very high). Hence, employing increments of 0.2 effectively provides a dependency score on a 0–1 scale.
[35] To assign ENCORE's production processes to each 4-digit NACE we employ the crosswalk of ISIC, NACE and GICS developed by UNEP-WCMC and SBTN. See https://sciencebasedtargetsnetwork.org/wp-content/uploads/2022/02/Sectoral-Materiality-Tool_UNEP-WCMC_January-2022.xlsx
[36] Using the highest vulnerability of each production process is justified by the low substitutability of ecosystem services. For example, agriculture processes have very large dependencies on water provision services; and a lack of water cannot be compensated (or averaged down) by a lower dependency on other ecosystem services such as maintenance of nursery habitat or animal-based energy.

By construction, the $NRS$ quantifies nature risks on a 0–1 scale, providing a simple metric that can be used to compare and evaluate firms' nature-related risks. In our risk scenario, we employ a horizon $T$ = 2050, thus considering the impact of nature deterioration up to 2050.

### 3.5 Firms' losses

We use two complementary methods to estimate firms' losses in a scenario of unabated nature decline[37].

**Stock market appraisal.** To evaluate firms' losses in equity markets, we employ a simple model where market returns are driven by both nature-related and financial risks. As nature degradation intensifies, firms with large nature-related exposures and vulnerabilities will be particularly impacted by the loss of nature and its services. Hence, we consider that firms' losses in an adverse scenario are correlated with nature risk, estimating stock market losses as:

$$Loss_{SM}^i = - NRS^i \cdot \sigma_{i,j}^m \cdot l_{i,j}^m \tag{10}$$

where $NRS^i$ is nature risk of each firm, whereas $\sigma_{i,j}^m$ and $l_{i,j}^m$ are the volatility and leverage multipliers. Beyond nature risk, firms' losses in an adverse scenario will be also affected by their financial risk profile. Therefore, we calculate the volatility and leverage multipliers as:

$$\sigma_{i,j}^m = \frac{\sigma_i}{\bar{\sigma}_j} \; ; \; l_{i,j}^m = \frac{l_i}{\bar{l}_j} \tag{11}$$

where $\sigma_i$ and $l_i$ are the volatility and financial leverage of firm $i$, while and $\bar{\sigma}_j$ and $\bar{l}_j$ are the sectoral averages. To discriminate among the best- and worst-in-class in each economic sector, we rely on the classification employed by the ECB and ESRB to assess climate risks in financial portfolios, employing 31 industries to assess financial risks[38].

**Discounted cash flow valuation.** To complement the stock market method, we also perform a DCF valuation where cash flows are affected by nature risk. If nature deterioration continues apace, firms with large nature-related risks will face diminished future cash flows, driven by declining revenues and/or increasing operating costs. Hence, we consider that cash flows are impacted by the $NRS$, obtaining firms' risk-adjusted values as:

$$\tilde{V}_{DCF}^i = \frac{\sum_{t=t0}^{T-1} CF_t^i (1 - NRS_t^i)}{(1 + WACC)^t} + \frac{CF_T^i (1 + g)(1 - NRS_T^i)}{(WACC - g)(1 + WACC)^T} \tag{12}$$

---

[37] We reckon that these methods are an oversimplification of the complex relation between nature risks and financial losses. However, time is short to wait for the perfect model (Trust et al., 2023). Hence, while acknowledging their limitations, we consider that these methods can provide reasonable initial estimates of the losses that different firms may face from nature degradation.

[38] Specifically, we first classify companies by NACE section. Next, following the ESRB and ECB classification, we separate the manufacture (NACE C) and transport (H) sections into subsections, obtaining 31 economic industries. Table 3 describes each economic industry. See also EIOPA (2022), ECB (2022) and Crisóstomo (2022).

The state of nature deterioration in eq. (12) is considered through a time-variant version of the Nature Risk Score, $NRS_t^i$, which is calculated yearly up to 2050. Consequently, nature risk is progressively considered in cash flows, increasing as the degradation of nature intensifies. To set the weighted average cost of capital, $WACC$, and the growth rate, $g$, we rely on Damodaran (2022, 2023).[39] Moreover, we assume that highly leveraged companies will incur larger losses in a scenario of financial distress, as they are particularly vulnerable to declining operating margins. Therefore, firms' losses are computed as the difference between each firm initial value, $V_0^i$, and the risk-adjusted value, $\tilde{V}_{DCF}^i$, and adjusted by firms' financial leverage, as reflected in eq. (13)[40]:

$$Loss_{DCF}^i = (\tilde{V}_{DCF}^i - V_0^i)\, l_{i,j}^m \qquad (13)$$

### 3.6 Firms' data

We employ the MSCI World Index to obtain a representative sample of global companies. For each firm, we retrieve all the information required in our nature risk framework: NACE level 4 code, ENCORE dependencies, geographical exposures, market volatility, financial leverage and country-level nature degradations. Our sample includes all firms with available data. The final database comprises 1454 companies, domiciled in 45 countries, and covers over 94% of the MSCI World capitalization. Table 2 summarizes the data retrieved for each company.

Table 2: Firm-level information, data sources and modeling use

| Firm-level data | Data sources | Modeling use |
| --- | --- | --- |
| NACE level 4 code | Refinitiv, Bloomberg | Nature-related vulnerabilities |
| Ecosystem dependencies | ENCORE | Nature-related vulnerabilities |
| Spatial revenues | Refinitiv, Bloomberg | Geographical exposures |
| Environmental data | NGFS, IUCN, UN's statistics, NCI | Nature degradation hazards |
| Market volatility | Refinitiv, Bloomberg | Financial risk |
| Financial leverage | Refinitiv, Bloomberg | Financial risk |

---

[39] The global $WACC$ for equities is set at 7.26%, which is the average cost of capital for global equities in 2022-2023 (Damodaran, 2022, 2023). The growth rate $g$ is heuristically set at 2.59% to provide an initial value $V_0^i$ of 100 for all firms in absence of nature risk ($NRS_t = 0$), using a $WACC$ of 7.26% and projected cash flows $CF_t = 5(1+g)^t$.

[40] We reckon that leverage is generally considered in DCF valuations through adjusted cost of capital and cash flows estimates. However, to obtain firm-level valuations that specifically discriminate the effect of nature risk, our framework employs cash flow that are adjusted by nature risk and a global $WACC$, including leverage through equation (13) by considering that highly leveraged firms ($l_{i,j}^m > 1$) will be particularly impacted in an adverse scenario of reduced operating margins.

# 4. Results

This section presents the results from our nature risk framework. First, we analyze the damages caused by environmental hazards in specific countries. Next, we consider the vulnerability of economic sectors to nature loss. Finally, we combine country-level nature degradations, sectoral vulnerabilities and firms' geographical exposures to quantify the risk that individual firms face due to the decline of nature and its services.

## 4.1 Geographical risk: Country Degradation Index

Figure 2 shows the nature deterioration expected in 174 countries in a scenario of unabated nature decline. We estimate that 53% of the world's nature could be damaged or impaired by 2050[41]. The loss of nature is widespread across all countries, with damages to ecosystems ranging from 14% to 79%. Consequently, the current path of nature overexploitation poses a huge threat to the environment, jeopardizing the essential goods and services that sustain human life.

**Figure 2: Nature degradation projected in 2050**

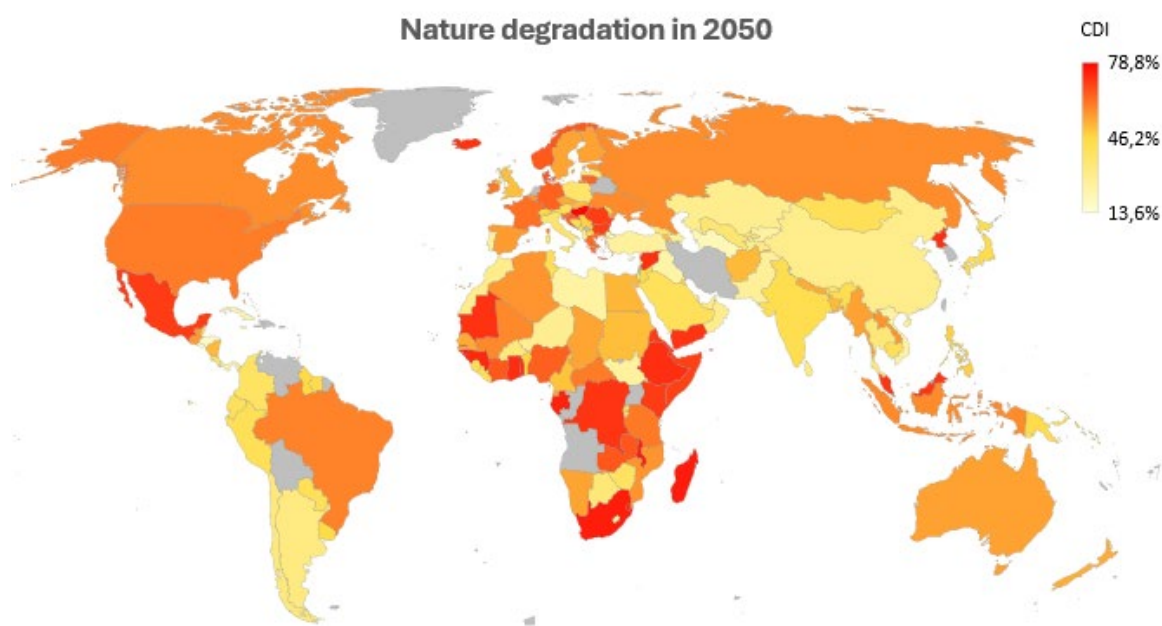

Notes: The *CDI* measures the impact of nature degradation in specific countries. It is calculated by aggregating five environmental hazards: biodiversity loss, land degradation, climate change, population growth and nature capital depletion; and considers both nonlinear effects and tipping points.

Countries in sub-Saharan Africa are the most affected by nature risk, with 24 out of 45 territories suffering damages to ecosystems greater than 60%. Nature loss in these territories is driven by soil degradation —53% decline in fertile soils by 2050—, climate change —~1°C temperature increase projected in 2023-2050— and nature capital depletion.

---

[41] The impact of nature degradation is assessed in 174 countries. To evaluate nature degradation in supranational regions (including world nature), we employ land-weighted averages of the deterioration expected in the countries within each region.

North America and Europe also suffer heightened nature loss. Global warming is expected to cause extensive environmental damage at higher latitudes, with temperatures rising 2.3°C above pre-industrial levels. In addition, North America is negatively impacted by nature capital depletion —52% decline by mid-century—, whereas acute land degradation is expected in some European countries (e.g., Slovenia, Romania and Ukraine).

Alternatively, nature loss in Eastern Asia, the Caribbean, and Pacific Island countries is comparatively lower. Nature risk in Eastern Asia is driven by the $CDI$ of China, where ecosystems are expected to decline by 30%. Environmental deterioration in China is lower than average due to encouraging trends in soil conservation and nature capital. In the Caribbean, nature-related damages are projected to reach 22%, mainly due to reduced soil erosion and biodiversity loss. Similarly, moderate global warming and land degradation are the main drivers of nature loss projections in Oceania, where the $CDI$ of small island countries predicts ecosystem damages of 37%[42]. Annex 1 shows the nature degradation projected for 174 countries representing 93.3% of the world's GDP.

Figure 3 shows the range of damages expected from different nature risk hazards. Overall, climate change and a growing human population are the main environmental pressures driving nature loss —with expected damages of 70% and 63%, on average, by mid-century. Similarly, nature capital and fertile lands also face significant declines, with average damages of 55% and 30%. However, the impacts from degraded soils and nature capital vary notably by region. Several countries are at risk of losing most of their fertile lands and natural resources by 2050 (e.g., South Africa, Mexico and Hungary). In contrast, other territories show encouraging data, improving soil quality (Panama and Sweden) or recovering nature capital (Cuba and Thailand).

**Figure 3: Range of damages expected from environmental risk hazards in 2050**

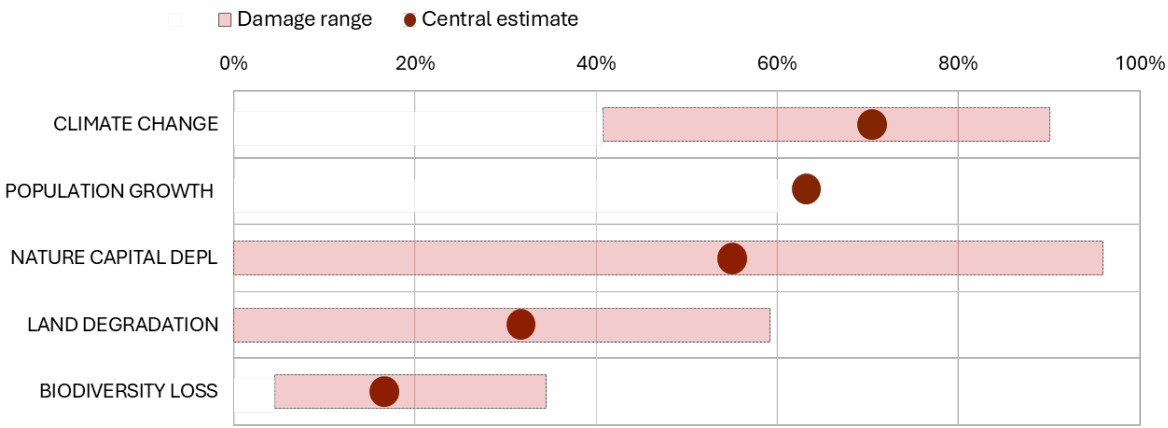

Notes: The range of damages shows the variability in damage projections in 2050. For nature capital and land degradation, given large country-level variabilities, the interquartile range is shown. Human population is modelled as a systemic risk factor driving a global degradation of nature across all regions.

---

[42] Small island countries exclude Australia, New Zealand and Papua New Guinea.

Comparatively, the rate of biodiversity loss is smaller than other risk hazards. However, even small biodiversity losses can have far-reaching consequences, altering the ecological balance that sustains environmental systems. Researchers consider that a safe limit for biodiversity might have been already breached[43]. Using IUCN data, our projections show biodiversity damages ranging from 4.5% to 34.4% in 2023-2050. Although lower than other hazards, incremental losses of an already damaged biodiversity pose a notable risk to the environment, threatening the complex biological systems that support nature's delivery of goods, services and amenities.

## 4.2 Sectoral risk: Vulnerability Score

Figure 4 shows the vulnerability of economic sectors to nature loss. For 81 NACE level 2 industries, nature's vulnerabilities range from 37.5% to 100%, as measured by the $VS$. Overall, agriculture firms show the highest vulnerability to nature loss (90.4% on average) owing to very large dependencies on water-related services, pest and disease control and soil quality. Similarly, the $VS$ of utilities, transport and mining is particularly large (80.9%, 78.6% and 76% respectively) due to high nature related dependencies on climate regulation and flood and storm control, among others.

**Figure 4: Vulnerability to nature by economic sector**

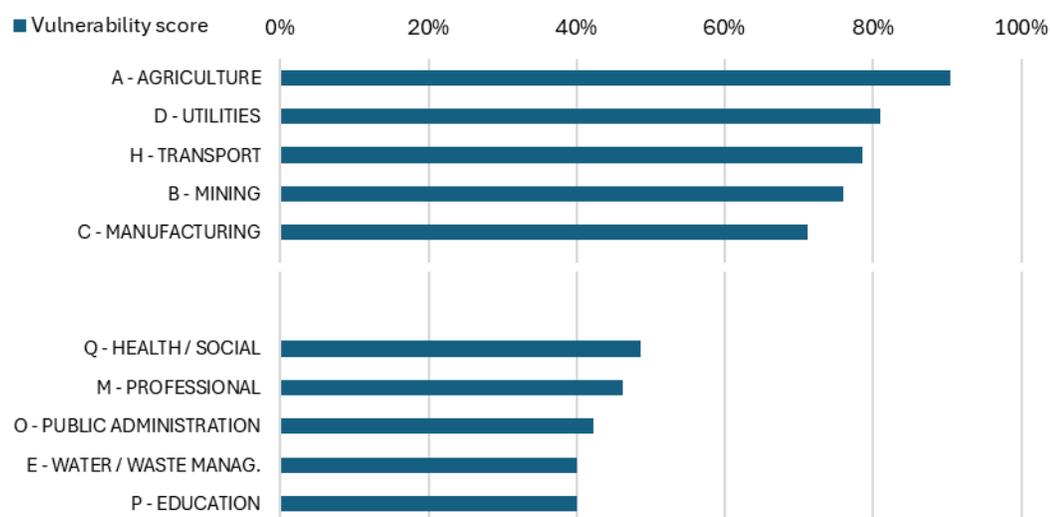

Source: Own calculations based on ENCORE

To discriminate firms' vulnerabilities to nature, we perform a granular assessment of 601 NACE level 4 segments. For several industries, we find significant differences in the vulnerabilities of specific activities compared to their sectoral aggregates. For instance, leasing of water transportation equipment has a vulnerability score of 100%, whereas the renting and leasing industry exhibits a $VS$ of 61.6%. Similarly, while the telecommunication industry shows very large ecosystem vulnerabilities (92.5% on average), satellite telecommunication exhibits a lower score of 70%.[44]

---

[43] See e.g., De Palma et al. (2021), Newbold et al. (2016) and Steffen et al. (2015).

Remarkably, even the least affected companies show significant nature-related dependencies, showing a vulnerability of ~40%. For instance, although firms in scientific research or sewerage are among the least exposed to the loss of ecosystem services, a decline in genetic material or biological filtration may significantly impact these companies, generating sizable nature-related risks.

Beyond direct vulnerabilities, unabated nature loss is also expected to cause widespread economic strains, generating disruptions in supply chains, labor productivity and resource availability. Moreover, continued loss of nature can fuel social unrest, increasing the risk of wars and mass migrations. Consequently, nature degradation poses a significant risk for all companies, even those without large direct nature dependencies. Annex 2 shows the vulnerability score and main ecosystem dependencies of a representative sample of economic activities.

**4.3 Firm-specific risk: Nature Risk Score**

The $NRS$ quantifies the risk that companies face from nature loss given their geographical exposures, sectoral vulnerabilities and the deterioration of nature expected in specific countries. Through a broad assessment of 174 countries, 601 economic activities and 1454 firms, we find that all MSCI World companies face significant nature-related risks. The average $NRS$ is 0.33, with risk scores ranging from 0.11 to 0.65 depending on firms' characteristics.[45]

Figure 5 shows that nature risk varies substantially both across and within economic sectors. Utilities (0.48 $NRS$) and real estate companies (0.43) show the highest average $NRS$, whereas health and social activities (0.25) and administrative services (0.27) are more resilient to nature loss[46]. However, there are significant intra-sector variabilities for individual companies. For instance, while the information sector has a moderate risk (0.27), wired telecommunications companies in Europe and North America exhibit a risk score of 0.60. Conversely, although mining and quarrying has a relatively high nature risk (0.41), one company extracting iron ores in China has a risk score of 0.19. These examples illustrate the importance of considering firm-level characteristics in nature risk assessment, showing that sectoral aggregates can misrepresent the risk of individual companies.

---

[44] Renting and leasing of water transport equipment (NACE 77.34) is particularly vulnerable to nature because it is directly related to the highly nature-dependent water transport activities. In contrast, satellite telecommunication (NACE 61.30) exhibits lower nature dependencies than the telecommunication industry because it is less exposed to natural disasters (e.g., floods or storms).

[45] We consider an equal-weighted average of the MSCI World components to avoid an overrepresentation of certain sectors and countries. A capitalization weighting would have resulted in a slightly lower average NRS (0.31).

[46] The financial services sector exhibits an average $NRS$ of 0.22, mainly due to comparatively lower ENCORE dependencies. However, ENCORE's assessment of financial services dependencies is primarily based on the vulnerabilities of the companies' office buildings and does not consider the credit or investment portfolio of financial companies, which could lead to notably higher nature-related vulnerabilities.

**Figure 5: Nature Risk Score by economic sector**

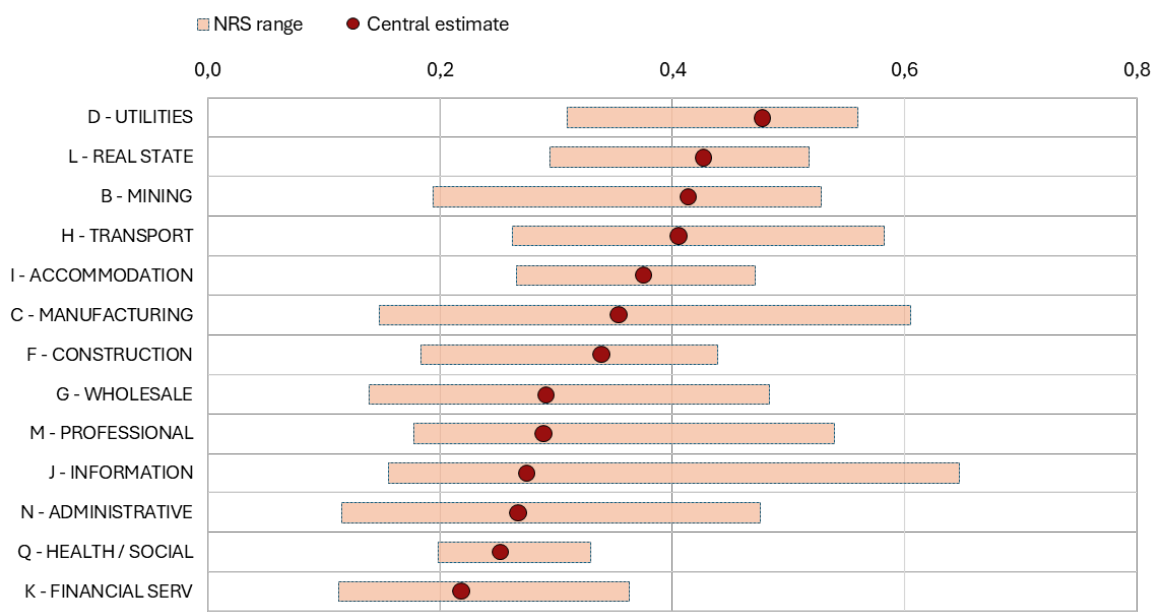

Notes: NACE sectors with more than 10 companies are shown.

As expected, nature risk is particularly acute in companies operating in countries facing severe environmental degradation and industries that are particularly vulnerable to nature loss. Besides wired telecommunication firms in Europe and North America, manufacturers of food products (0.54 $NRS$) and water transport companies (0.51 $NRS$) are heavily affected by nature risk. In contrast, firms in advertising and education are among the most resilient to nature loss, with a $NRS$ of 0.21 and 0.19, on average.

### 4.4 Firm-level losses from nature risk

Figure 6 shows the loss of value projected for 1454 companies in a scenario of unabated nature decline. We find that global equities suffer a 26.8% loss, on average, due to the decline of nature and its services. However, the loss distribution is significantly left-skewed, with the 1% worst-performing firms shedding 74.2% of their value. Firms' losses are estimated by aggregating the expected impact from our stock market appraisal and discounted cash flow methods[47].

---

[47] Expected losses are obtained by averaging the impact from our stock market and DCF methods. The stock market method considers that nature deterioration projections for 2023-2050 are reflected in stock prices at time $t = t_0$. Alternatively, the DCF method considers that nature deterioration is reflected in cash flows as it materializes, providing a lower severity estimate of potential firm losses (see Figure 6). By aggregating both methods, our loss estimates assume that nature degradation is partially anticipated by financial markets, while some of the effects are only incorporated as they materialize. This is consistent with how financial markets price novel risks (see e.g., Brunnermeier et al., 2021; Eren et al., 2022 and Karagozoglu 2021).

**Figure 6: MSCI World firms' losses from nature degradation**

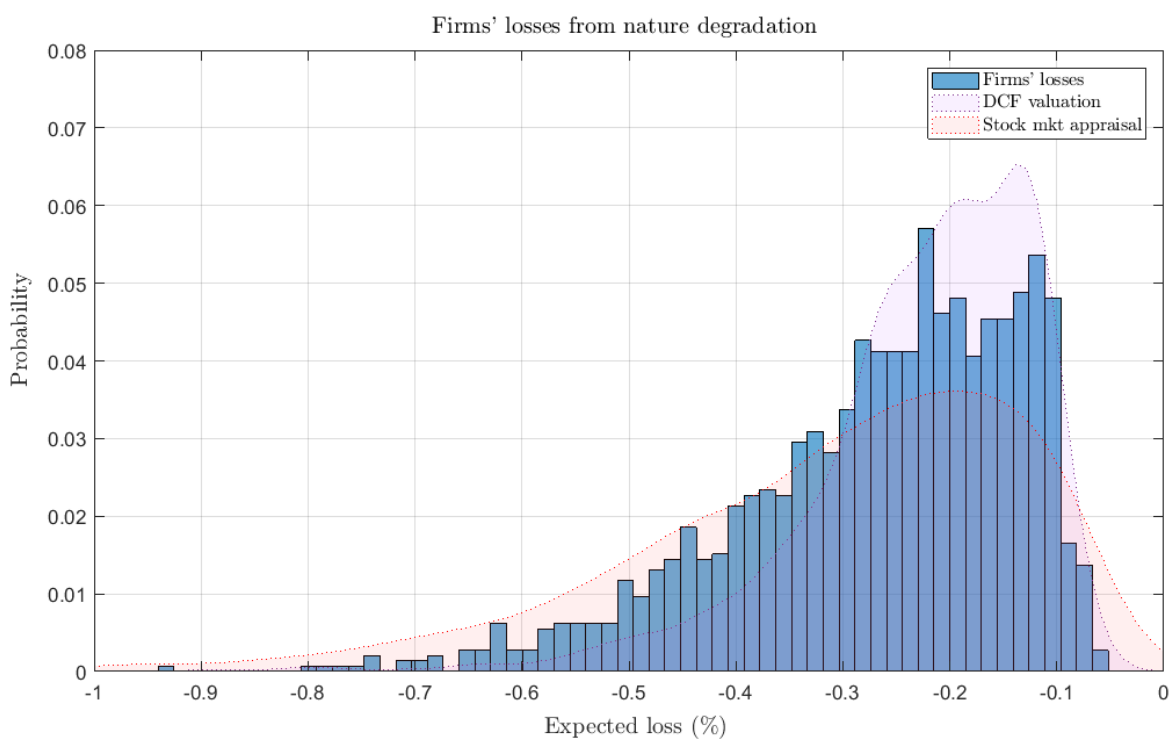

Our estimates suggest that all economic sectors will suffer substantial losses if nature degradation continues apace. For 31 different industries, firms' losses vary from 15% to 43% (see Table 3). Manufacture of food and beverages (-43.1%), water transportation (-40.9%) and utilities (-38.5%) face the highest losses. In contrast, firms in water and waste management (-17.2%) and education (-14.9%) show lower impacts from nature risk.

We also find that firm-level characteristics may significantly alter the risks that companies face from nature deterioration. Beyond sectoral differences, the 1% worst-performing companies (-74.2%) are highly leveraged and volatile firms (73% debt to asset and 45% annual volatility) with exposures concentrated in territories facing severe degradation (e.g., North America, Europe and Japan.)

Alternatively, the best-performing firms are geographically diversified companies with low financial risk (5.3% debt to assets and 18.4% volatility). However, even these firms suffer a sizable 7.4% loss. Consequently, our analyses suggest that nature degradation poses a stark risk even for the most resilient companies, showing that all businesses face large-scale losses if we do not halt the degradation of nature and its services.

**Table 3: Expected loss by sector in a scenario of unabated nature decline**

| NACE | Equity prices (% change) | Industry description |
|---|---|---|
| A01-03 | -37.09% | Crop, animal production, hunting and related services |
| B05-B09 | -32.78% | Mining and quarrying |
| C10-C12 | -43.12% | Manufacture of food products, beverages and tobacco |
| C13-C18 | -32.61% | Manufacture of textiles, wearing apparel, leather, paper and related products |
| C19 | -26.45% | Manufacture of coke and refined petroleum products |
| C20 | -29.95% | Manufacture of chemicals and chemical products |
| C21-C22 | -35.68% | Manufacture of pharmaceutical products and preparations, rubber and plastic products |
| C23 | -36.84% | Manufacture of other non-metallic mineral products |
| C24-C25 | -28.35% | Manufacture of basic and fabricated metal products, except machinery and equipment |
| C26-C28 | -22.86% | Manufacture of computer, electronic, optical, electrical equipment and machinery |
| C29-C30 | -25.70% | Manufacture of motor vehicles, trailers and semi-trailers and other transport equipment |
| C31-33 | -28.23% | Manufacture of furniture, repair and installation of machinery and other manufacturing |
| D35 | -38.49% | Electricity, gas, steam and air conditioning supply |
| E36-E39 | -17.26% | Water supply; sewerage; waste management and remediation activities |
| F41-F43 | -27.78% | Construction |
| G45-G47 | -22.40% | Wholesale and retail trade; repair of motor vehicles and motorcycles |
| H49 | -29.18% | Land transport and transport via pipelines |
| H50 | -40.87% | Water transport |
| H51 | -32.65% | Air transport |
| H52-H53 | -34.19% | Warehousing and support activities for transportation; Postal and courier activities |
| I55-I56 | -29.16% | Accommodation and Food Service Activities |
| J58-J63 | -22.68% | Information and Communication |
| K64-K66 | -18.04% | Financial and Insurance Activities |
| L68 | -34.69% | Real Estate Activities |
| M69-M75 | -22.82% | Professional, Scientific and Technical Activities |
| N77-N82 | -24.43% | Administrative and Support Service Activities |
| O84 | -21.16% | Public Administration and Defense; Compulsory Social Security |
| P85 | -14.87% | Education |
| Q86-Q88 | -20.91% | Human Health and Social Work Activities |
| R90-R93 | -36.16% | Arts, Entertainment and Recreation |
| S94-S96 | -18.76% | Other Service Activities |

# 5. Conclusion

This paper provides a comprehensive framework to quantify the financial risks that firms face due to nature deterioration. Combining country-level projections of nature degradation, sectoral vulnerabilities and firms' geographical exposures, we find that nature loss poses a stark threat to our environment and global economy, jeopardizing the essential goods and services that sustain human life.

Using five environmental metrics —biodiversity, climate, human population, nature capital and fertile soils—, we estimate that 53% of the world's nature could be damaged or impaired by 2050. The loss of nature is widespread across all countries, with ecosystem damages ranging from 14% to 79%. The impact of nature deterioration is particularly acute in Sub-Saharan Africa, Europe and North America.

We find that global equities could suffer a 26.8% loss, on average, if nature degradation continues apace. However, the nature risk distribution is significantly skewed, with the worst-performing companies losing ~75% of their value. Across different industries, the losses from nature overexploitation vary from 15% to 43%.

Firms in agriculture, mining and transport are particularly vulnerable to nature loss due to high dependencies on water-related services, climate regulation and access to raw materials, among others. Moreover, we find that firm-level characteristics can significantly alter the risk that companies face from nature deterioration. Through several examples, we show that sectoral averages can potentially misrepresent the risk of specific companies, leading to biases in nature risk assessments.

Looking forward, we outline several directions for future research. First, since nature degradation is driven by a myriad of complex interrelated systems, further work is needed to better understand how environmental pressures interact and how to measure their joint impact. Similarly, future research could further explore the nuanced links between nature degradation and financial risk. Overall, by providing a comprehensive framework to quantify nature-related risks, we aim to shed light on the losses that companies may face from nature deterioration, and encourage businesses, policy makers and researchers to think about the future consequences of nature loss.

# 6. References

Acheampong, A. O., & Opoku, E. E. O. (2023). Environmental degradation and economic growth: Investigating linkages and potential pathways. *Energy Economics*, *123*. https://doi.org/10.1016/j.eneco.2023.106734

Armstrong McKay, D. I., Abrams, J. F., Winkelmann, R., Sakschewski, B., Loriani, S., Fetzer, I., Cornell, S. E., Rockström, J., & Lenton, T. M. (2022). Exceeding 1.5°C global warming could trigger multiple climate tipping points. *Science*, *377*(6611). https://doi.org/10.1126/science.abn7950

Boldrini, S., Ceglar, A., Lelli, C., Parisi, L., & Heemskerk, I. (2023). Living in a World of Disappearing Nature: Physical Risk and the Implications for Financial Stability. *ECB Occasional Paper Series*, *333*. https://doi.org/10.2139/ssrn.4630721

Botero, C. A., Weissing, F. J., Wright, J., & Rubenstein, D. R. (2015). Evolutionary tipping points in the capacity to adapt to environmental change. *Proceedings of the National Academy of Sciences of the United States of America*, *112*(1). https://doi.org/10.1073/pnas.1408589111

Brunnermeier, M., Farhi, E., Koijen, R. S. J., Krishnamurthy, A., Ludvigson, S. C., Lustig, H., Nagel, S., & Piazzesi, M. (2021). Review Article: Perspectives on the Future of Asset Pricing. In *Review of Financial Studies* (Vol. 34, Issue 4). https://doi.org/10.1093/rfs/hhaa129

Calice, P., Kalan, F. D., & Miguel, F. (2021). Nature-related financial risks in Brazil. *Policy Research Working Paper - World Bank Group*. https://openknowledge.worldbank.org/bitstream/handle/10986/36201/Nature-Related-Financial-Risks-in-Brazil.pdf?sequence=1&isAllowed=y

CISL. (2020). Biodiversity Loss and Land Degradation An Overview of the Financial Materiality. *University of Cambridge Institute for Sustainability Leadership*.

Costanza, R., de Groot, R., Sutton, P., van der Ploeg, S., Anderson, S. J., Kubiszewski, I., Farber, S., & Turner, R. K. (2014). Changes in the global value of ecosystem services. *Global Environmental Change*, *26*(1). https://doi.org/10.1016/j.gloenvcha.2014.04.002

Crisóstomo, R. (2022). Measuring Transition Risk in Investment Funds. *CNMV Working Papers No. 81*. https://doi.org/10.2139/ssrn.4252801

Damodaran, A. (2022). *Costs of Capital by Industry*. https://pages.stern.nyu.edu/~adamodar/

Damodaran, A. (2023). *Costs of Capital by Industry*. https://pages.stern.nyu.edu/~adamodar/

Dasgupta, P. (2021). The economics of biodiversity: the Dasgupta review. *HM Treasury*. https://doi.org/10.2458/jpe.2289

De Palma, A., Hoskins, A., Gonzalez, R. E., Börger, L., Newbold, T., Sanchez-Ortiz, K., Ferrier, S., & Purvis, A. (2021). Annual changes in the Biodiversity Intactness Index in tropical and subtropical forest biomes, 2001–2012. *Scientific Reports*, *11*(1). https://doi.org/10.1038/s41598-021-98811-1

Dietz, S., Rising, J., Stoerk, T., & Wagner, G. (2021). Economic impacts of tipping points in the climate system. *Proceedings of the National Academy of Sciences of the United States of America*, *118*(34). https://doi.org/10.1073/pnas.2103081118

ECB. (2022). *2022 Climate risk stress test*.

ECB/ESRB. (2023). *Towards macroprudential frameworks for managing climate risk*. December.

Eddy, T. D., Lam, V. W. Y., Reygondeau, G., Cisneros-Montemayor, A. M., Greer, K., Palomares, M. L. D., Bruno, J. F., Ota, Y., & Cheung, W. W. L. (2021). Global decline in capacity of coral reefs to provide ecosystem services. *One Earth*, *4*(9). https://doi.org/10.1016/j.oneear.2021.08.016

EIOPA. (2022). *2022 IORP stress test technical specifications*. April.

Eren, E., Merten, F., & Verhoeven, N. (2022). Pricing of climate risks in financial markets a summary of the literature. In *Bank for International Settlements*.

Estes, J. A., Terborgh, J., Brashares, J. S., Power, M. E., Berger, J., Bond, W. J., Carpenter, S. R., Essington, T. E., Holt, R. D., Jackson, J. B. C., Marquis, R. J., Oksanen, L., Oksanen, T., Paine, R. T., Pikitch, E. K., Ripple, W. J., Sandin, S. A., Scheffer, M., Schoener, T. W., … Wardle, D. A. (2011). Trophic downgrading of planet earth. In *Science* (Vol. 333, Issue 6040). https://doi.org/10.1126/science.1205106

Gibbs, H. K., & Salmon, J. M. (2015). Mapping the world's degraded lands. In *Applied Geography* (Vol. 57). https://doi.org/10.1016/j.apgeog.2014.11.024

Hillebrand, H., Donohue, I., Harpole, W. S., Hodapp, D., Kucera, M., Lewandowska, A. M., Merder, J., Montoya, J. M., & Freund, J. A. (2020). Thresholds for ecological responses to global change do not emerge from empirical data. *Nature Ecology and Evolution*, *4*(11). https://doi.org/10.1038/s41559-020-1256-9

IPBES. (2019). *Global assessment report of the Intergovernmental Science-Policy Platform on Biodiversity and Ecosystem Services*. https://ipbes.net/global-assessment%0Ahttps://ipbes.net/global-assessment-report-biodiversity-ecosystem-services

IPBES. (2022). *The Diverse values and valuation of nature - Summary for policymakers*.

IPCC. (2018). *Global Warming of 1.5 °C*.

IPCC. (2019). Climate Change and Land: an IPCC special report. In *Climate Change and Land: an IPCC Special Report on climate change, desertification, land degradation, sustainable land management, food security, and greenhouse gas fluxes in terrestrial ecosystems*.

IPCC. (2023). SYNTHESIS REPORT OF THE IPCC SIXTH ASSESSMENT REPORT (AR6). *Intergovernmental Panel on Climate Change*. https://doi.org/10.1017/9781009157896.015

Karagozoglu, A. K. (2021). Novel Risks: A Research and Policy Overview. In *Journal of Portfolio Management* (Vol. 47, Issue 9). https://doi.org/10.3905/JPM.2021.1.287

Kedward, K., Ryan-Collins, J., & Chenet, H. (2020). Managing Nature-Related Financial Risks: A Precautionary Policy Approach for Central Banks and Financial Supervisors. *UCL Institute for Innovation and Public Purpose, Working Paper Series*. https://doi.org/10.2139/ssrn.3726637

Keen, S. (2023). *Loading the DICE against Pension Funds: Flawed economic thinking on climate has put your pension at risk*. July.

Lenton, T. M., Rockström, J., Gaffney, O., Rahmstorf, S., Richardson, K., Steffen, W., & Schellnhuber, H. J. (2019). Climate tipping points — too risky to bet against. In *Nature* (Vol. 575, Issue 7784). https://doi.org/10.1038/d41586-019-03595-0

Lu, F., & Sohail, M. T. (2022). Exploring the Effects of Natural Capital Depletion and Natural Disasters on Happiness and Human Wellbeing: A Study in China. *Frontiers in Psychology*, *13*. https://doi.org/10.3389/fpsyg.2022.870623

Newbold, T., Hudson, L. N., Arnell, A. P., Contu, S., De Palma, A., Ferrier, S., Hill, S. L. L., Hoskins, A. J., Lysenko, I., Phillips, H. R. P., Burton, V. J., Chng, C. W. T., Emerson, S., Gao, D., Hale, G. P., Hutton, J., Jung, M., Sanchez-Ortiz, K., Simmons, B. I., … Purvis, A. (2016). Has land use pushed terrestrial biodiversity beyond the planetary boundary? A global assessment. *Science*, *353*(6296). https://doi.org/10.1126/science.aaf2201

Newbold, T., Tittensor, D. P., Harfoot, M. B. J., Scharlemann, J. P. W., & Purves, D. W. (2020). Non-linear changes in modelled terrestrial ecosystems subjected to perturbations. *Scientific Reports*, *10*(1). https://doi.org/10.1038/s41598-020-70960-9

NGFS. (2023). *Nature-related Financial Risks: a Conceptual Framework to guide Action by Central Banks and Supervisors Foreword 2*. *September*.

NGFS-INSPIRE. (2021). Biodiversity and financial stability: building the case for action. *NGFS Occasional Paper*, *October*.

OECD. (2022). Climate Tipping Points: Insights for Effective Policy Action. *OECD Publishing*. https://doi.org/10.1038/scientificamerican1113-6

OWID. (2023). *OWID based on Maddison Project Database 2020, GCIP and van Zanden et al. (2014)*. https://ourworldindata.org/history-of-poverty-data-appendix

Pörtner, H. O., Scholes, R. J., Arneth, A., Barnes, D. K. A., Burrows, M. T., Diamond, S. E., Duarte, C. M., Kiessling, W., Leadley, P., Managi, S., McElwee, P., Midgley, G., Ngo, H. T., Obura, D., Pascual, U., Sankaran, M., Shin, Y. J., & Val, A. L. (2023). Overcoming the coupled climate and biodiversity crises and their societal impacts. In *Science* (Vol. 380, Issue 6642). https://doi.org/10.1126/science.abl4881

Prince, S. D., Von Maltitz, G., Zhang, F., Byrne, K., Driscoll, C., Eshel, G., Kust, G., Martínez-Garza, C., Metzger, J. P., Midgley, G., Thwin, S., Moreno-Mateos, D., & Sghaier, M. (2018). Status and trends of land degradation and restoration and associated changes in biodiversity and ecosystem functions. In *The IPBES assessment report on land degradation and restoration* (Issue 2018).

Ranger, C., Alvarez, N., Freeman, J., Harwood, A., Obersteiner, T., Paulus, M., & Sabuco, E. (2023). The Green Scorpion: the Macro-Criticality of Nature for Finance Foundations for scenario-based analysis of complex and cascading physical nature-related financial risks force on biodiversity loss and nature-related risks (Task force Nature) of the Network. *NGFS Occasional Paper*, *December*. https://www.ngfs.net/en/the-green-scorpion-macro-criticality-nature-for-finance

Richardson, K., Steffen, W., Lucht, W., Bendtsen, J., Cornell, S. E., Donges, J. F., Drüke, M., Fetzer, I., Bala, G., von Bloh, W., Feulner, G., Fiedler, S., Gerten, D., Gleeson, T., Hofmann, M., Huiskamp, W., Kummu, M., Mohan, C., Nogués-Bravo, D., … Rockström, J. (2023). Earth beyond six of nine planetary boundaries. *Science Advances*, *9*(37). https://doi.org/10.1126/sciadv.adh2458

Rockström, J., Steffen, W., Noone, K., Persson, Å., Chapin, F. S., Lambin, E. F., Lenton, T. M., Scheffer, M., Folke, C., Schellnhuber, H. J., Nykvist, B., De Wit, C. A., Hughes, T., Van Der Leeuw, S., Rodhe, H., Sörlin, S., Snyder, P. K., Costanza, R., Svedin, U., … Foley, J. A. (2009). A safe operating space for humanity. In *Nature* (Vol. 461, Issue 7263). https://doi.org/10.1038/461472a

Scheffer, M., Carpenter, S., Foley, J. A., Folke, C., & Walker, B. (2001). Catastrophic shifts in ecosystems. In *Nature* (Vol. 413, Issue 6856). https://doi.org/10.1038/35098000

Steffen, W., Richardson, K., Rockström, J., Cornell, S. E., Fetzer, I., Bennett, E. M., Biggs, R., Carpenter, S. R., De Vries, W., De Wit, C. A., Folke, C., Gerten, D., Heinke, J., Mace, G. M., Persson, L. M., Ramanathan, V., Reyers, B., & Sörlin, S. (2015). Planetary boundaries: Guiding human development on a changing planet. *Science, 347*(6223). https://doi.org/10.1126/science.1259855

Svartzman, R., Espagne, E., Julien, G., Paul, H.-L., Mathilde, S., Allen, T., Berger, J., Calas, J., Godin, A., & Vallier, A. (2022). A "Silent Spring" for the Financial System? Exploring Biodiversity-Related Financial Risks in France. *SSRN Electronic Journal*. https://doi.org/10.2139/ssrn.4028442

Trust, S., Joshi, S., Lenton, T., & Oliver, J. (2023). The Emperor's New Climate Scenarios Limitations and assumptions of commonly used climate-change scenarios in financial services. *IFoA*, *July*. www.actuaries.org.uk

Turner, R. K., & Daily, G. C. (2008). The ecosystem services framework and natural capital conservation. *Environmental and Resource Economics*, *39*(1). https://doi.org/10.1007/s10640-007-9176-6

UNEP. (2021). Ecosystem restoration for people, nature and climate. In *Ecosystem restoration for people, nature and climate*. https://doi.org/10.4060/cb4927en

van der Esch, S., ten Brink, B., Stehfest, E., Bakkenes, M., Sewell, A., Bouwman, A., Meijer, J., Westhoek, H., van den Berg, M., & van den Born, G.J. Doelman, J. (2017). Exploring future changes on food, water, climate condition and the impacts in land use and land change and biodiversity: Scenarios for the UNCCD Global Land Outlook. In *PBL Netherlands Environmental Assessment Agency* (Issue 1).

van Toor, J., Piljic, D., Schellekens, G., Van Oorschot, M., & Kok, M. (2020). *Indebted to nature - Exploring biodiversity risks for the Dutch financial sector*. June, 44. https://www.dnb.nl/media/cy2p51gx/biodiversity-opportunities-risks-for-the-financial-sector.pdf

Volk, A. A., & Atkinson, J. A. (2013). Infant and child death in the human environment of evolutionary adaptation. *Evolution and Human Behavior*, *34*(3). https://doi.org/10.1016/j.evolhumbehav.2012.11.007

WEF. (2020). Nature Risk Rising: Why the Crisis Engulfing Nature Matters for Business and the Economy. *New Nature Economy Series*.

WEF. (2024). The Global Risks Report 2024. In *Angewandte Chemie International Edition, 6(11), 951–952.* (Vol. 2).

World Bank. (2022). An Exploration of Nature-Related Financial Risks in Malaysia. *An Exploration of Nature-Related Financial Risks in Malaysia*, March. https://doi.org/10.1596/37314

Xu, Y., & Ramanathan, V. (2017). Well below 2 °C: Mitigation strategies for avoiding dangerous to catastrophic climate changes. In *Proceedings of the National Academy of Sciences of the United States of America* (Vol. 114, Issue 39). https://doi.org/10.1073/pnas.1618481114

Zijdeman, R. L., & de Silva, F. R. (2014). Life expectancy since 1820. In *How Was Life?* https://doi.org/10.1787/9789264214262-10-en

# Annex 1 - Nature loss projection for 174 countries in 2050

Table 4: Nature degradation expected in individual countries

|   | Country | CDI |   | Country | CDI |   | Country | CDI |   | Country | CDI |
|---|---|---|---|---|---|---|---|---|---|---|---|
| 1 | Afghanistan | 51.3% | 45 | Dominica | 33.4% | 89 | Liechtenstein | 53.4% | 133 | Sao Tome-Princ. | 70.4% |
| 2 | Albania | 27.8% | 46 | Ecuador | 40.8% | 90 | Lithuania | 64.3% | 134 | Saudi Arabia | 40.8% |
| 3 | Algeria | 56.8% | 47 | Egypt | 51.7% | 91 | Luxembourg | 64.5% | 135 | Senegal | 53.5% |
| 4 | Andorra | 60.3% | 48 | El Salvador | 16.2% | 92 | Madagascar | 75.3% | 136 | Serbia | 47.3% |
| 5 | Antigua-Barbuda | 42.4% | 49 | Equatorial Guinea | 54.7% | 93 | Malawi | 75.1% | 137 | Seychelles | 26.4% |
| 6 | Argentina | 32.2% | 50 | Eritrea | 70.4% | 94 | Malaysia | 71.3% | 138 | Sierra Leone | 42.1% |
| 7 | Armenia | 50.8% | 51 | Estonia | 56.2% | 95 | Mali | 58.6% | 139 | Singapore | 40.4% |
| 8 | Australia | 54.9% | 52 | Eswatini | 75.0% | 96 | Malta | 37.9% | 140 | Slovakia | 50.9% |
| 9 | Austria | 41.8% | 53 | Ethiopia | 72.6% | 97 | Mauritania | 72.0% | 141 | Slovenia | 62.9% |
| 10 | Azerbaijan | 27.2% | 54 | Finland | 54.7% | 98 | Mauritius | 20.9% | 142 | Solomon Islands | 37.1% |
| 11 | Bahamas | 52.2% | 55 | France | 62.4% | 99 | Mexico | 70.4% | 143 | Somalia | 69.7% |
| 12 | Bahrain | 27.0% | 56 | Gabon | 74.5% | 100 | Mongolia | 41.5% | 144 | South Africa | 75.1% |
| 13 | Bangladesh | 51.2% | 57 | Gambia | 75.4% | 101 | Montenegro | 38.9% | 145 | South Sudan | 25.6% |
| 14 | Belgium | 56.1% | 58 | Georgia | 35.3% | 102 | Morocco | 29.8% | 146 | Spain | 55.8% |
| 15 | Belize | 47.6% | 59 | Germany | 64.7% | 103 | Mozambique | 56.9% | 147 | Sri Lanka | 31.8% |
| 16 | Benin | 47.1% | 60 | Ghana | 72.8% | 104 | Myanmar | 54.8% | 148 | State of Palestine | 66.4% |
| 17 | Bhutan | 38.2% | 61 | Greece | 63.5% | 105 | Namibia | 54.5% | 149 | Sudan | 50.6% |
| 18 | Bolivia | 46.6% | 62 | Guatemala | 56.2% | 106 | Nepal | 53.3% | 150 | Suriname | 44.3% |
| 19 | Bosnia-Herzegov. | 43.5% | 63 | Guinea | 73.5% | 107 | Netherlands | 44.5% | 151 | Sweden | 54.5% |
| 20 | Botswana | 34.1% | 64 | Guinea-Bissau | 75.3% | 108 | New Zealand | 53.2% | 152 | Switzerland | 45.8% |
| 21 | Brazil | 59.5% | 65 | Guyana | 45.9% | 109 | Nicaragua | 51.2% | 153 | Syria | 72.4% |
| 22 | Bulgaria | 70.1% | 66 | Honduras | 17.2% | 110 | Niger | 28.7% | 154 | Tajikistan | 31.0% |
| 23 | Burkina Faso | 33.3% | 67 | Hungary | 78.8% | 111 | Nigeria | 64.9% | 155 | Thailand | 34.3% |
| 24 | Burundi | 37.8% | 68 | Iceland | 71.8% | 112 | North Macedonia | 44.7% | 156 | Timor-Leste | 60.2% |
| 25 | Cabo Verde | 55.7% | 69 | India | 43.2% | 113 | Norway | 66.1% | 157 | Togo | 73.9% |
| 26 | Cambodia | 41.6% | 70 | Indonesia | 58.3% | 114 | Oman | 29.9% | 158 | Tonga | 20.1% |
| 27 | Cameroon | 49.7% | 71 | Iran | 74.0% | 115 | Pakistan | 27.0% | 159 | Trinidad-Tobago | 38.5% |
| 28 | Canada | 58.1% | 72 | Iraq | 23.7% | 116 | Palau | 44.2% | 160 | Tunisia | 42.4% |
| 29 | Central African R. | 60.6% | 73 | Ireland | 61.5% | 117 | Panama | 25.4% | 161 | Türkiye | 26.0% |
| 30 | Chad | 54.4% | 74 | Israel | 28.9% | 118 | Papua New Guinea | 44.3% | 162 | Turkmenistan | 19.8% |
| 31 | Chile | 33.4% | 75 | Italy | 37.5% | 119 | Paraguay | 41.0% | 163 | Ukraine | 58.3% |
| 32 | China | 29.9% | 76 | Jamaica | 13.6% | 120 | Peru | 41.8% | 164 | UAE | 32.9% |
| 33 | Colombia | 38.8% | 77 | Japan | 44.3% | 121 | Philippines | 47.9% | 165 | United Kingdom | 50.2% |
| 34 | Congo | 71.5% | 78 | Jordan | 45.4% | 122 | Poland | 37.6% | 166 | Tanzania | 60.9% |
| 35 | Costa Rica | 37.5% | 79 | Kazakhstan | 29.3% | 123 | Portugal | 25.9% | 167 | USA | 60.5% |
| 36 | Côte d'Ivoire | 66.1% | 80 | Kenya | 70.4% | 124 | Qatar | 36.2% | 168 | Uruguay | 45.5% |
| 37 | Croatia | 66.1% | 81 | Kuwait | 44.2% | 125 | Moldova | 45.8% | 169 | Uzbekistan | 35.8% |
| 38 | Cuba | 20.9% | 82 | Kyrgyzstan | 25.1% | 126 | Romania | 70.3% | 170 | Venezuela | 53.4% |
| 39 | Cyprus | 18.3% | 83 | Lao People's | 56.9% | 127 | Russia | 58.1% | 171 | Viet Nam | 24.8% |
| 40 | Czechia | 53.7% | 84 | Latvia | 35.3% | 128 | Rwanda | 41.9% | 172 | Yemen | 71.7% |
| 41 | Korea | 72.2% | 85 | Lebanon | 23.6% | 129 | St Kitts and Nev. | 44.0% | 173 | Zambia | 66.3% |
| 42 | Congo | 62.8% | 86 | Lesotho | 25.6% | 130 | Saint Lucia | 42.4% | 174 | Zimbabwe | 39.2% |
| 43 | Denmark | 67.4% | 87 | Liberia | 41.8% | 131 | St Vincent-Gren. | 40.9% |   |   |   |
| 44 | Djibouti | 72,6% | 88 | Libya | 24,7% | 132 | San Marino | 61,2% |   |   |   |

Notes: The *CDI* measures the impact of nature degradation in specific countries. It is calculated by aggregating five environmental hazards: biodiversity loss, land degradation, climate change, population growth and nature capital depletion; and considers both nonlinear effects and tipping points.

# Annex 2 - Vulnerability score, production processes and ecosystem dependencies

Table 5: Vulnerability score and ecosystem dependencies of different economic activities

| NACE | Economic activity | Industry | ENCORE production processes | Vulnerability Score | Main ecosystem dependencies |
|---|---|---|---|---|---|
| 01.47 | Raising of poultry | A. Agriculture | Large-scale livestock (beef and dairy)<br>Small-scale livestock (beef and dairy) | 100% | Fibres and other materials; disease control<br>Flood and storm protection; soil quality<br>Surface & ground water; water quality<br>Nursery and habitat maintenance<br>Climate regulation |
| 08.11 | Quarrying of ornamental and building stone, limestone, gypsum, chalk and slate | B. Mining | Construction materials production | 100% | Surface & ground water |
| 10.83 | Processing of tea and coffee | C. Manufacturing | Processed food and drink production | 100% | Surface & ground water |
| 23.51 | Manufacture of cement | C. Manufacturing | Construction materials production | 100% | Surface & ground water |
| 61.10 | Wired telecommunications activities | J. Information | Telecommunication and wireless services | 100% | Flood and storm protection |
| 77.34 | Renting and leasing of water transport equipment | N. Administrative | Marine transportation | 100% | Climate regulation<br>Flood and storm protection |
| 35.11 | Production of electricity | D. Utilities | Biomass & geothermal energy production<br>Hydropower production<br>Infrastructure holdings<br>Wind & solar energy provision<br>Nuclear and thermal power stations<br>Electric & nuclear power transmission and distribution | 92.5% | Climate regulation<br>Flood and storm protection<br>Water flow maintenance<br>Surface water<br>Mass stabilisation and erosion control |
| 01.12 | Growing of rice | A. Agriculture | Large-scale irrigated arable crops<br>Large-scale rainfed arable crops<br>Small-scale irrigated arable crops<br>Small-scale rainfed arable crop | 80.0% | Flood and storm protection; soil quality<br>Mass stabilization and erosion control,<br>Surface & ground water; water quality<br>Water flow maintenance; climate regulation |
| 68.10 | Buying and selling of own real estate | L. Real State | Real estate activities | 80.0% | Surface & ground water |
| 14.11 | Manufacture of leather clothes | C. Manufacturing | Natural fibre production<br>Synthetic fibre production<br>Jewellery production | 73.3% | Fibres and other materials<br>Flood and storm protection<br>Water flow maintenance<br>Surface & ground water |
| 73.11 | Advertising agencies | M. Professional | Infrastructure holdings | 40.0% | Mass stabilisation and erosion control |